\documentclass[letterpaper,twocolumn,english,showpacs,preprintnumbers,amsmath,amssymb,superscriptaddress,nofootinbib,prl]{revtex4}
\usepackage[T1]{fontenc}
\usepackage[latin9]{inputenc}
\usepackage{float}
\usepackage{amsmath}
\usepackage{graphicx}
\usepackage{amssymb}

\makeatletter

\@ifundefined{textcolor}{}
{%
 \definecolor{BLACK}{gray}{0}
 \definecolor{WHITE}{gray}{1}
 \definecolor{RED}{rgb}{1,0,0}
 \definecolor{GREEN}{rgb}{0,1,0}
 \definecolor{BLUE}{rgb}{0,0,1}
 \definecolor{CYAN}{cmyk}{1,0,0,0}
 \definecolor{MAGENTA}{cmyk}{0,1,0,0}
 \definecolor{YELLOW}{cmyk}{0,0,1,0}
 }

\makeatother

\usepackage{babel}

\begin{document}

\title{Density of States of Quantum Spin Systems from Isotropic Entanglement}

\author{Ramis Movassagh}

\email[Corresponding author: ]{ramis@mit.edu }

\affiliation{Department of Mathematics, Massachusetts Institute of Technology,
Cambridge, MA, 02139}

\author{Alan Edelman}

\affiliation{Department of Mathematics, Massachusetts Institute of Technology,
Cambridge, MA, 02139}

\date{\today}

\pacs{TBF}

\begin{abstract}
We propose a method which we call \textbf{{}``Isotropic Entanglement''}
(IE), that predicts the eigenvalue distribution of quantum many body
(spin) systems (QMBS) with generic interactions. We interpolate between
two known approximations by matching fourth moments. Though, such
problems can be QMA-complete, our examples show that IE provides an
accurate picture of the spectra well beyond what one expects from
the first four moments alone.\textit{ }We further show that the interpolation
is universal, i.e., independent of the choice of local terms.
\end{abstract}
\maketitle

\section{\textit{\label{sec:Intro}}\textup{A highly accurate match (better
than four moments).}}

We propose a method to compute the {}``density of states'' (DOS)
or {}``eigenvalue density'' of quantum spin systems with generic
local interactions \cite{IE}. More generally one wishes to compute
the DOS of the sum of non-commuting random matrices from their, individually
known, DOS's. 

\begin{figure}
\centering{}\includegraphics[width=8cm]{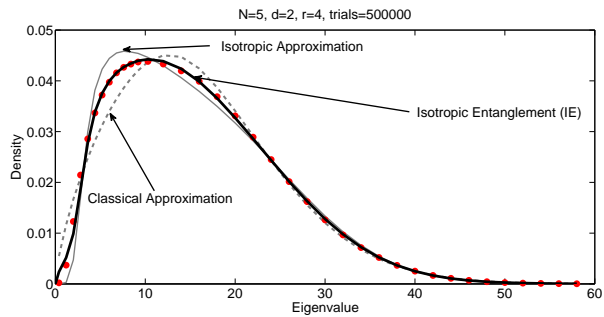}\caption{\label{fig:The_intro}The exact diagonalization in dots and IE compared
to the two approximations. The title parameters are explained in the
section on numerical results.}
\end{figure}

We begin with an example in Figure \ref{fig:The_intro}, where we
compare exact diagonalization against two approximations that we considered
early in our work and our method \cite{IE}. 
\begin{itemize}
\item Dashed grey curve: \textit{classical} approximation. Notice that it
overshoots to the right.
\item Solid grey curve: \textit{isotropic} approximation (or \textit{iso}).
Notice that it overshoots to the left.
\item Solid black curve: \textit{isotropic entanglement (IE).}
\item Dots: \textit{exact diagonalization} of the quantum problem given
in Eq. \ref{eq:Hamiltonian}.
\end{itemize}

The \textit{classical approximation} ignores eigenvector structure
by summing random eigenvalues uniformly from non-commuting matrices.
The curve is the convolution of the probability densities of the eigenvalues
of each matrix. 

The\textit{ isotropic approximation} assumes that the eigenvectors
are in {}``general position''; that is, we add the two matrices
with correct eigenvalue densities but choose the eigenvectors from
Haar (uniform) measure. As the matrix size goes to infinity, the resulting
distribution has been popularized recently as the free convolution
of the individual distributions \cite{IE}, or \cite[for a nice exposition]{speicher}. 

The exact diagonalization given by red dots, the dashed and solid
grey curves have exactly the same first three moments, but differing
fourth moments. 

\textit{Isotropic Entanglement (IE)} is a linear combination of the
two approximations obtained by matching the fourth moments. We show
1) the fit is better than what might be expected by the first four
moments alone, 2) the combination is always convex for the problems
of interest, given by a $0\leq p\leq1$, and 3) this convex combination
is universal depending on the parameter counts of the problem but
not the eigenvalue densities of the local terms.

\textit{\label{sec:Parameter-Counts}Parameter counts: exponential,
polynomial and zero.---}Because of the \textit{locality} of generic
interactions, the complete set of eigenstates has parameter count
equal to a polynomial in the number of spins, though the dimensionality
is exponential. The classical and isotropic approximations have zero
and exponentially many random parameters respectively. This suggests
that the problem of interest somehow lies between the two approximations. 

\vspace{-0.2in}

\section{\label{sec:Physical-Implications-and}\textup{Physical implications
and problem formulation.}}

QMBS spectra have been elusive for two reasons: 1. The terms that
represent the interactions are generally non-commuting. This is particularly
pronounced for systems with random interactions (e.g., quantum spin
glasses \cite[p. 320]{sachdev}\cite{spinGlass,sachdev2}). 2. Standard
numerical diagonalization is limited by memory and computer speed
because of exponential growth. Energy eigenvalue distributions are
needed for calculating the partition function $Z$. Exact calculation
of the spectrum of interacting QMBS has been shown to be difficult
\cite{schuch}. 

Though, much progress has been made in understanding the ground states
of interacting QMBS \cite{cirac,vidal,white,wen,frank,hastings,ramis},
eigenvalue distributions are less studied. An accurate description
of tails of distributions are desirable for condensed matter physics.
IE provides a direct method for obtaining eigenvalue distributions
of quantum spin systems with generic local interactions (i.e., quantum
spin systems) and does remarkably well at approximating the tails. 

Though we are not restricted to one dimensional chains, for sake of
concreteness, we investigate $N$ interacting $d$-dimensional quantum
spins (qudits) on a line with generic interactions. The Hamiltonian
is

\vspace{-0.15in}

\begin{equation}
H=\sum_{l=1}^{N-1}\mathbb{I}_{d^{l-1}}\otimes H_{l,\cdots,l+L-1}\otimes\mathbb{I}_{d^{N-l-\left(L-1\right)}},\label{eq:Hamiltonian}\end{equation}

\vspace{-0.15in}

\begin{flushleft}
where the local terms $H_{l,\cdots,l+L-1}$ are finite $d^{L}\times d^{L}$
random matrices. From now on we take the case of nearest neighbors
interactions, $L=2$, unless otherwise specified. 
\par\end{flushleft}

The eigenvalue distribution of any commuting subset of $H$ such as
the terms with $l$ odd (the {}``odds'') or $l$ even (the {}``evens'')
can be obtained using local diagonalization. However, the difficulty
in approximating the full spectrum of $H\equiv H_{\mbox{odd}\vphantom{\mbox{even}}}+H_{\mbox{even}\vphantom{\mbox{odd}}}$
is in summing the odds and the evens because of their overlap at every
site. 

The intuition behind IE is that terms with an overlap, such as $H_{l,l+1}$
and $H_{l+1,l+2}$, introduce randomness and mixing through sharing
of a site. Namely, the process of entanglement generation introduces
an \textit{isotropicity }between the eigenvectors of evens and odds
that can be harnessed to capture the spectrum.

\vspace{-0.1in}

\section{\label{sec:Isotropic-Entanglement}\textup{Isotropic entanglement.}}

Let $A$ and $B$ denote diagonal matrices containing the eigenvalues,
in random order, of the odds and evens respectively. In a basis where
(at least) the odds are diagonal, our approximations to the Hamiltonian
in Eq. \ref{eq:Hamiltonian} become

\vspace{-0.2in}

\begin{align}
\mbox{Classical: } & H^{c}=A+B\nonumber \\
\mbox{Isotropic: } & H^{iso}=A+Q^{-1}BQ,\label{eq:approximations of H}\end{align}

\vspace{-0.2in}

\begin{flushleft}
where $Q$ is a $\beta-$Haar measure orthogonal matrix of size $d^{N}$
with $\beta=1$ corresponding to real orthogonals, $\beta=2$ unitaries,
and $\beta=4$ quaternions (see \cite{alanGhost} for formal treatment
of a general $\beta$). 
\par\end{flushleft}

The Hamiltonian given by Eq. \ref{eq:Hamiltonian} in a basis where
the odds are diagonal is

\vspace{-0.2in}

\begin{flushleft}
\begin{equation}
\mbox{Quantum: }H=A+Q_{q}^{-1}BQ_{q}.\label{eq:TheProblem}\end{equation}

\par\end{flushleft}

\vspace{-0.2in}

\begin{flushleft}
The $Q_{q}$ has the formidable form $Q_{q}=\left(Q_{A}\right)^{-1}Q_{B}$,
where (for odd number of sites $N$)
\par\end{flushleft}

\vspace{-0.4in}

\begin{align}
Q_{A}= & Q_{1}\otimes Q_{3}\otimes\cdots\otimes Q_{N-2}\otimes\mathbb{I}_{d}\nonumber \\
Q_{B}= & \mathbb{I}_{d}\otimes Q_{2}\otimes Q_{4}\otimes\cdots\otimes Q_{N-1}.\label{eq:oddEvenQ}\end{align}

\vspace{-0.2in}

\begin{flushleft}
The eigenvalue distribution of Eq. \ref{eq:TheProblem} is what we
are after. 
\par\end{flushleft}

\begin{flushleft}
To establish the plausibility of the classical and isotropic approximations,
we consider the moments. In general we have the $k^{\textrm{th}}$
moments\begin{align}
m_{k}^{c}= & \frac{1}{d^{N}}\mathbb{E}\textrm{Tr}\left(A+B\right)^{k},\nonumber \\
m_{k}^{iso}= & \frac{1}{d^{N}}\mathbb{E}\textrm{Tr}\left(A+Q^{-1}BQ\right)^{k}\;\mbox{ and}\label{eq:moments}\\
m_{k}^{q}= & \frac{1}{d^{N}}\mathbb{E}\textrm{Tr}\left(A+Q_{q}^{-1}BQ_{q}\right)^{k}.\nonumber \end{align}

\par\end{flushleft}

\begin{flushleft}
The first three moments are usually encoded as the mean, variance,
and skewness; the fourth moment is encoded by the kurtosis which we
denote $\gamma_{2}$.
\par\end{flushleft}

\textbf{\label{thm:(The-Matching-Three}(The Matching Three Moments
Theorem) }The first three moments of the classical and isotropic approximations
exactly match those of the quantum problem \cite{IE}. 

Turning to the fourth moment, we propose to match the kurtosis $\gamma_{2}^{q}$
with a linear combination of the classical ($\gamma_{2}^{c}$) and
isotropic ($\gamma_{2}^{iso}$) kurtoses:

\vspace{-0.2in}

\begin{equation}
\gamma_{2}^{q}=p\gamma_{2}^{c}+\left(1-p\right)\gamma_{2}^{iso}\Rightarrow\qquad p=\frac{\gamma_{2}^{q}-\gamma_{2}^{iso}}{\gamma_{2}^{c}-\gamma_{2}^{iso}}.\label{eq:convex}\end{equation}

\begin{flushleft}
In terms of probability measures, IE provides
\par\end{flushleft}

\vspace{-0.2in}

\begin{equation}
d\nu^{q}\approx d\nu^{IE}=pd\nu^{c}+(1-p)d\nu^{iso},\label{eq:measureConvex}\end{equation}

\vspace{-0.2in}

\begin{flushleft}
where $d\nu$ denotes a probability measure. So long as $0\leq p\leq1$,
$d\nu^{IE}$ is a probability measure whose first four moments match
the theoretical measure $d\nu^{q}$.
\par\end{flushleft}

In the expansion of the fourth moments, by a theorem we call The Departure
Theorem\cite{IE}, the numerator and the denominator in Eq. \ref{eq:convex}
respectively become,

\begin{align}
\gamma_{2}^{q}-\gamma_{2}^{iso}= & \frac{2}{d^{N}}\frac{\mathbb{E}\left\{ \textrm{Tr}\left[\left(AQ_{q}^{-1}BQ_{q}\right)^{2}-\left(AQ^{-1}BQ\right)^{2}\right]\right\} }{\sigma^{4}}\nonumber \\
\gamma_{2}^{c}-\gamma_{2}^{iso}= & \frac{2}{d^{N}}\frac{\mathbb{E}\left\{ \textrm{Tr}\left[\left(AB\right)^{2}-\left(AQ^{-1}BQ\right)^{2}\right]\right\} }{\sigma^{4}}.\label{eq:numer}\end{align}

Evaluation of $p$ in Eq. \ref{eq:convex} reduces to the evaluation
of the right hand sides of Eq. \ref{eq:numer}. It is natural to ask,
does a $0\leq p\leq1$ always exist such that the convex combination
Eq. \ref{eq:convex} can be formed? How would quantum spectra look
in the thermodynamical limit ($N\rightarrow\infty$)? The Slider Theorem
provides the answers. 

\textbf{(The Slider Theorem)\label{thm:The-Slider-Theorem} }The quantum
kurtosis lies in between the classical and the isotropic kurtoses,
$\gamma_{2}^{iso}\leq\gamma_{2}^{q}\leq\gamma_{2}^{c}$. Therefore
there exists a $0\leq p\leq1$ such that $\gamma_{2}^{q}=p\gamma_{2}^{c}+\left(1-p\right)\gamma_{2}^{iso}$.
Furthermore, $\lim_{N\rightarrow\infty}p=1$.

\begin{figure}[H]
\begin{centering}
\includegraphics[width=8cm]{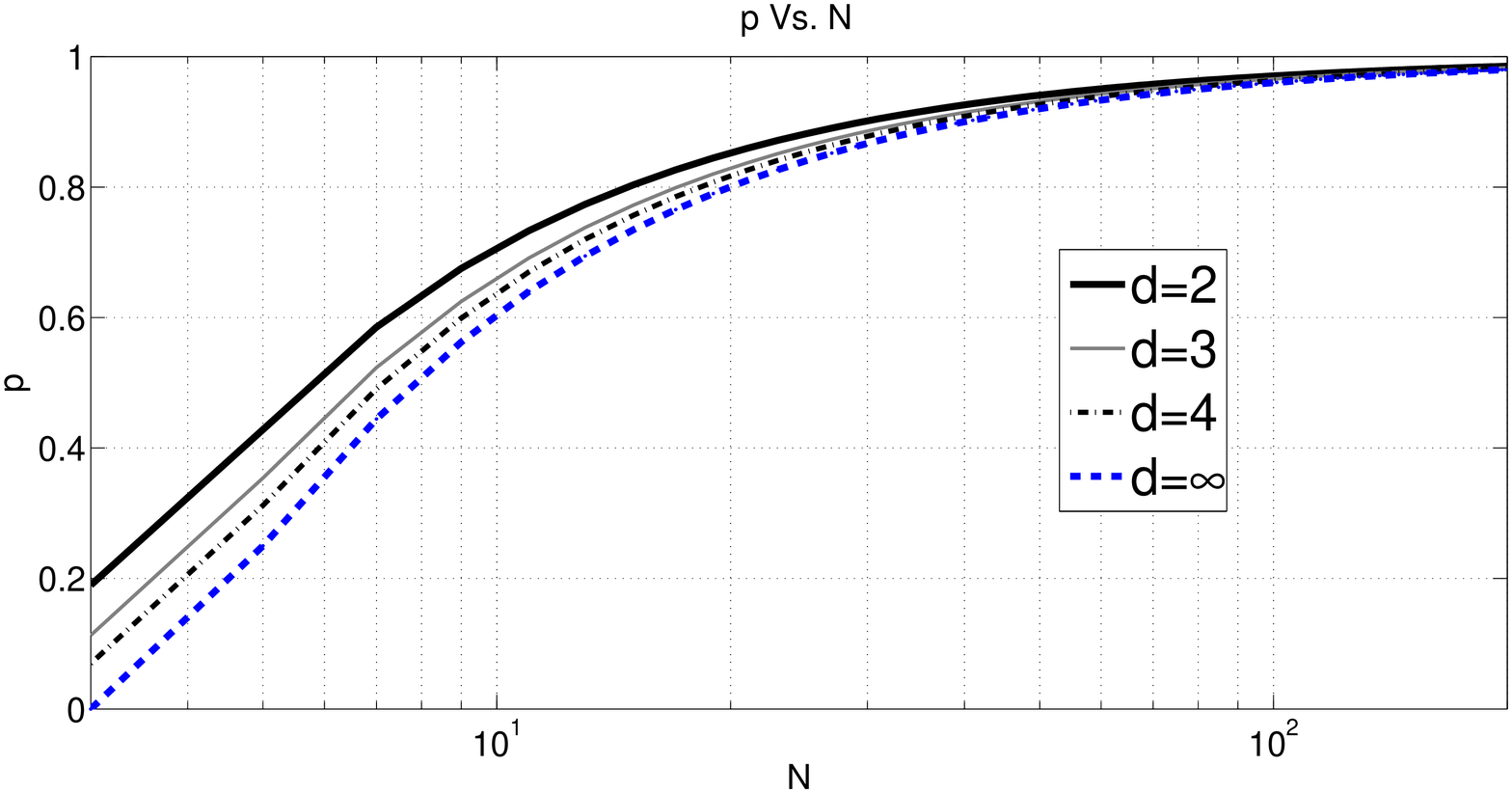}
\par\end{centering}

\centering{}\caption{\label{fig:pVsN}An example where $\beta=1$: the quantum problem
for all $d$ lies in between the iso $(p=0)$ and the classical $(p=1)$. }
\end{figure}

Through our numerical investigations, we observed that $p$ did not
change by the choice of local terms; it did not dependent on the covariance
matrix. This lead us to prove that $p$ was universal with respect
to the choice of generic local terms and allowed us to derive an analytical
formula (Figure \ref{fig:pVsN}, where we take $\beta=1$).

\textbf{(Universality Corollary)} $p\mapsto p\left(N,d,\beta\right)$,
namely, it is independent of the distribution of the local terms (see
\cite{IE} for an analytical formula).

\vspace{-0.3in}

\vspace{-0.3in}

\begin{flushleft}
Entanglement shows itself starting at the fourth moment; further,
in the expansion of the fourth moments, only the terms that involve
\textit{a pair} of local terms \textit{sharing a site} differ \cite{IE}. 
\par\end{flushleft}

\vspace{-0.3in}

\begin{figure}[H]
\centering{}\includegraphics[width=8cm]{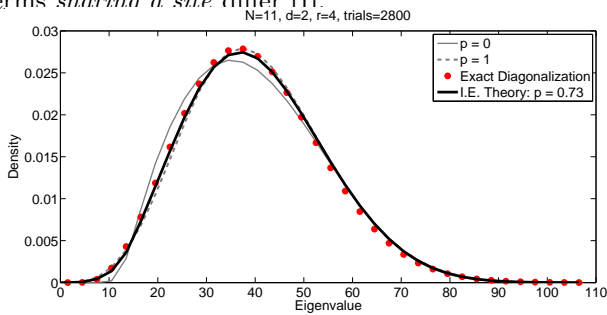}\caption{\label{fig:N=00003D11}$N=11$ with full rank Wishart matrices as
local terms.}
\end{figure}

\begin{figure}[H]
\begin{centering}
\includegraphics[width=8cm]{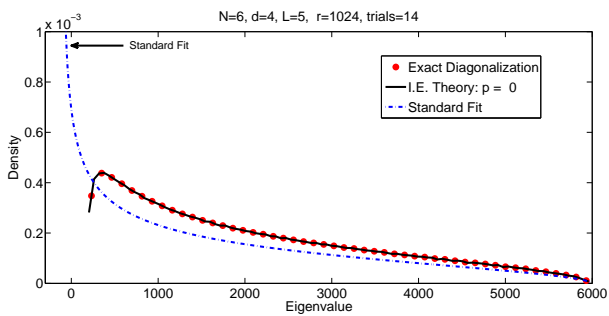}
\par\end{centering}

\caption{\label{fig:L=00003D5}Approximating the quantum spectrum by $H^{IE}=\sum_{l=1}^{2}Q_{l}^{T}H_{l,\cdots,l+4}Q_{l}$,
where the interaction is taken to be $5-$local Wishart matrices. }
\end{figure}

\vspace{-0.2in}

\section{\label{sec:Numerical-Results}Numerical results.}

Here we compare our theory against exact diagonalization for various
number of sites $N$, local ranks $r$, and site dimensionality $d$.
As our first two examples (Figures \ref{fig:The_intro} and \ref{fig:N=00003D11})
we take the local terms to be independent Wishart matrices: Each $H_{l,l+1}=A^{T}A$,
where $A$ is a $d^{2}\times d^{2}$ matrix with independent and identically
distributed Gaussian random entries. The higher moments of the Wishart
matrices were obtained using MOPs \cite{mops}.

One would expect that $L>2$ would be closer to the isotropic approximation
than when the interactions are nearest neighbors because of the larger
number of random parameters in Eq. \ref{eq:Hamiltonian}. When the
number of random parameters of the problem $\left(N-L+1\right)d^{L}$
becomes comparable to $d^{N}$, we indeed find that we can approximate
the spectrum with a high accuracy by taking the summands to be all
isotropic \cite{ramisPI} (Figure \ref{fig:L=00003D5}). 

\begin{figure}
\begin{centering}
\includegraphics[width=8cm]{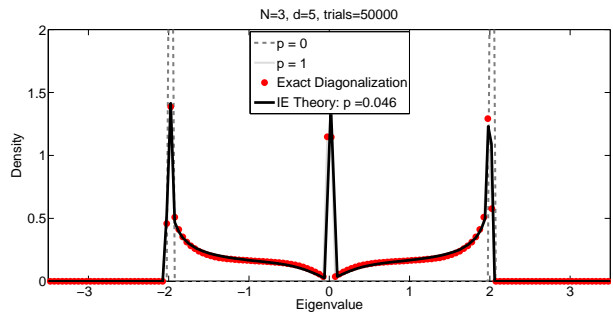}
\par\end{centering}

\caption{\label{fig:binomial}Local terms have a random binomial distribution.}
\end{figure}

Lastly take the local terms to have Haar eigenvectors but with Bernoulli
eigenvalues: $H_{l,l+1}=Q_{l}^{T}\Lambda_{l}Q_{l}$, where $\Lambda_{l}$
is a diagonal matrix of random eigenvalues $\pm1$ (Figure \ref{fig:binomial}).
Here classical treatment leads to a binomial distribution. As expected
$p=1$ in Figure \ref{fig:binomial} has three atoms at $-2,0,2$
corresponding to the randomized sum of the eigenvalues from the two
local terms. The exact diagonalization, however, shows that the quantum
chain has a much richer structure closer to iso; i.e, $p=0$. This
is captured quite well by IE with $p=0.046$. 

\begin{figure}
\raggedright{}\includegraphics[scale=0.32]{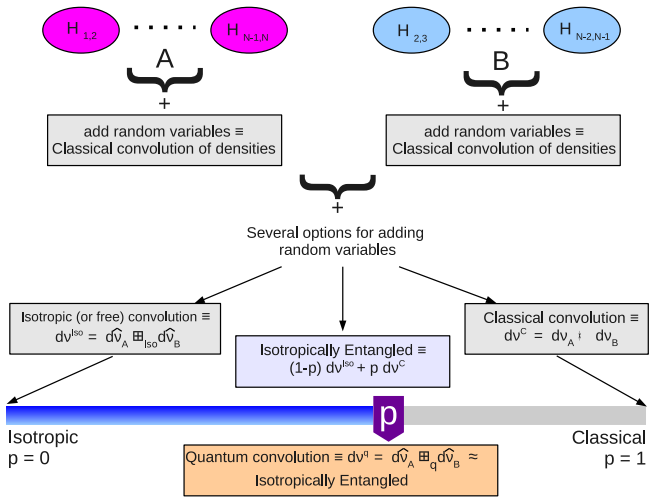}\caption{\label{fig:Summary}Summary of Isotropic Entanglement. We have denoted
the known classical and isotropic convolutions by $\star$ and $\boxplus_{\mbox{iso}}$
respectively. IE approximates the unknown quantum convolutions $\boxplus_{q}$. }
\end{figure}

Most distributions built solely from the first four moments would
give smooth curves. Roughly speaking, the mean indicates the center
of the distribution, variance its width, skewness its bending away
from the center and kurtosis how tall and skinny versus how short
and fat the distribution is. It is hard to imagine that the kinks,
cusps and local extrema of the quantum problem (as seen in some of
our examples and in particular Figure \ref{fig:binomial}) could be
captured by fitting only the first four moments of the QMBS Hamiltonian
to a known distribution. Remarkably a one parameter (i.e., $p$) interpolation
between the isotropic and classical suffices in capturing the richness
of the spectra of QMBS. Figure \ref{fig:Summary} summarizes IE.

\vspace{-0.2in}

\section{\label{sec:Accuracy-Beyond-Four}Accuracy beyond four moments.}

\begin{figure}
\begin{centering}
\includegraphics[width=8cm]{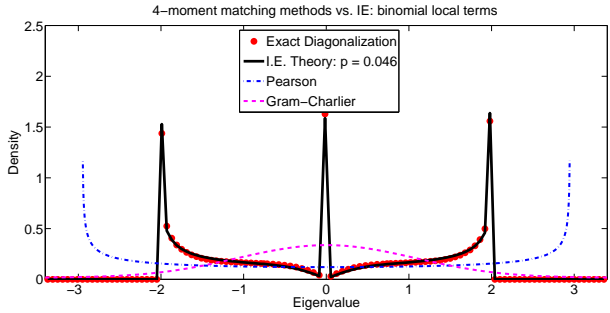}
\par\end{centering}

\begin{centering}
\caption{\label{fig:IEvsOthersZERO}IE vs Pearson and Gram-Charlier fits for
local terms with binomial distribution}

\par\end{centering}

\end{figure}

\begin{figure}
\begin{centering}
\includegraphics[width=8cm]{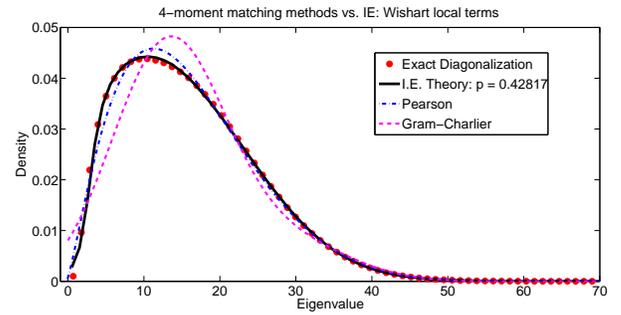}\caption{\label{fig:IEvsOthersWishart}IE vs Pearson and Gram-Charlier fits
for local terms with Wishart distribution}

\par\end{centering}

\end{figure}

We illustrate in Figures \ref{fig:IEvsOthersZERO} and \ref{fig:IEvsOthersWishart}
how the IE fit is better than expected when matching four moments.
We used the exact first four moments to approximate the density using
the Pearson fit as implemented in MATLAB and also the well-known Gram-Charlier
fit \cite{Cramer}. In \cite{Gram} it was demonstrated that the statistical
mechanics methods for obtaining the DOS, when applied to a finite
dimensional vector space, lead to a Gaussian distribution in the lowest
order. Further, they showed that successive approximations lead naturally
to the Gram-Charlier series \cite{Cramer}. Comparing these against
the accuracy of IE leads us to view IE as more than a moment matching
methodology. 

\vspace{-0.2in}

\section{\textup{\label{sec:Outlook}Outlook.}}

\vspace{-0.1in}

Our work supports a very general principle that one can obtain an
accurate representation of inherently exponential problems dealing
with QMBS by approximating them with far less complexity. This realization
is at the heart of other recent developments in QMBS research such
as Matrix Product States \cite{cirac,vidal}, and Density Matrix Renormalization
Group \cite{white}, where the \textit{state} (usually the ground
state of $1D$ chains) can be adequately represented by a Matrix Product
State (MPS) ansatz whose parameters grow \textit{linearly} with the
number of quantum particles. Future work includes explicit treatment
of fermionic systems and numerical exploration of higher dimensional
systems.

We thank Peter W. Shor, Jeffrey Goldstone, Patrick Lee, Peter Young,
Gil Strang, Mehran Kardar, Xiao-Gang Wen and Ofer Zeitouni for discussions.
We thank the National Science Foundation for support through grants
CCF-0829421 and DMS-1035400.


\vspace{-0.2in}

\begin{thebibliography}{18}
\bibitem{IE}R. Movassagh, A. Edelman, arXiv:1012.5039 (2010)

\bibitem{speicher}A. Nica, R. Speicher, Lectures on the Combinatorics
of Free Probability (Cambridge University Press 2006)

\bibitem{sachdev}S. Sachdev, Quantum Phase Transitions (Cambridge
University Press 2001)

\bibitem{sachdev2}S. Sachdev Physics World 7, No. 10, 25 (Oct. 1994)

\bibitem{spinGlass}M. Fannes, B. Nachtergaele, and R. F.Werner, Commun.
Math. Phys. 144, 443 (1992)

\bibitem{schuch}B. Brown, S. T. Flammia, N. Schuch, Phys. Rev. Lett.
107, 040501 (2011)

\bibitem{hastings}A. Ambainis, A. W. Harrow, M B. Hastings, arXiv:0910.0472v2
(2010)

\bibitem{cirac}D. Perez-Garcia, F. Verstraete, M. M. Wolf, and J.
I. Cirac, Quantum Inf. Comput. 7, 401 (2007)

\bibitem{vidal}G. Vidal, Phys. Rev. Lett. 93, 040502 (2004). See
\cite{cirac} for an alternative proof.

\bibitem{white}S. R. White, Phys. Rev. Lett. 69, 2863 (1992)

\bibitem{wen}Z. C. Gu, M. Levin, and X. G. Wen, Phys. Rev. B 78,
205116 (2008)

\bibitem{frank}F. Verstraete, V. Murg, and J. I. Cirac, Adv. Phys.
57, 143 (2008)

\bibitem{ramis}R. Movassagh, E. Farhi, J. Goldstone, D. Nagaj, T.
J. Osborne, P. W. Shor, Phys. Rev. A 82, 012318 (2010)

\bibitem{alanGhost} A. Edelman, pp. 783-790. Volume 16 Issue 4 (2010)

\bibitem{mops}I. Dumitriu, A. Edelman and G. Shuman, Journal of Symbolic
Computation 42, 587-620 (2007)

\bibitem{ramisPI}R. Movassagh, A. Edelman, Talk given at The Perimeter
Institute (July 2010): {\footnotesize pirsa.org/index.php?p=speaker\&name=Ramis\_Movassagh}{\footnotesize \par}

\bibitem{Cramer}H. Cramér, Mathematical Methods of Statistics, (Princeton
University Press, Princeton 1957)

\bibitem{Gram}S. Das Guptaa, R. K. Bhaduri, Physics Letters B, vol
58 issue 4, 381-383 (1975)
\end{thebibliography}
\end{document}